\documentclass[12pt]{article}
\usepackage{ringberg}

\makeatletter
\let\hangafter\@hangfrom

\newcount\@tempcntc
\def\@citex[#1]#2{\if@filesw\immediate\write\@auxout{\string\citation{#2}}\fi
  \@tempcnta\z@\@tempcntb\m@ne\def\@citea{}\@cite{\@for\@citeb:=#2\do
    {\@ifundefined
       {b@\@citeb}{\@citeo\@tempcntb\m@ne\@citea\def\@citea{,}{\bf ?}\@warning
       {Citation `\@citeb' on page \thepage \space undefined}}%
    {\setbox\z@\hbox{\global\@tempcntc0\csname b@\@citeb\endcsname\relax}%
     \ifnum\@tempcntc=\z@ \@citeo\@tempcntb\m@ne
       \@citea\def\@citea{,}\hbox{\csname b@\@citeb\endcsname}%
     \else
      \advance\@tempcntb\@ne
      \ifnum\@tempcntb=\@tempcntc
      \else\advance\@tempcntb\m@ne\@citeo
      \@tempcnta\@tempcntc\@tempcntb\@tempcntc\fi\fi}}\@citeo}{#1}}
\def\@citeo{\ifnum\@tempcnta>\@tempcntb\else\@citea\def\@citea{,}%
  \ifnum\@tempcnta=\@tempcntb\the\@tempcnta\else
   {\advance\@tempcnta\@ne\ifnum\@tempcnta=\@tempcntb \else \def\@citea{--}\fi
    \advance\@tempcnta\m@ne\the\@tempcnta\@citea\the\@tempcntb}\fi\fi}
\makeatother
\renewcommand{\section}[1] {\vspace*{0.6cm}\addtocounter{sectionc}{1}
 \setcounter{subsectionc}{0}\setcounter{subsubsectionc}{0}\noindent
 {\normalsize\bf #1}\par\vspace*{0.4cm}}
\renewcommand{\subsection}[1] {\vspace*{0.6cm}\addtocounter{subsectionc}{1}
 \setcounter{subsubsectionc}{0}\noindent
 {\normalsize\it #1}\par\vspace*{0.4cm}}
\def\lsim{\mathrel{\raise.3ex\hbox{$<$\kern-.75em\lower1ex\hbox{$\sim$}}}}
\def\gsim{\mathrel{\raise.3ex\hbox{$>$\kern-.75em\lower1ex\hbox{$\sim$}}}}
\def\lplm{\ell^+\ell^-}
\def\mpmm{\mu^+\mu^-}
\def\epem{e^+e^-}
\def\9{\hphantom0}
\def\ifmath#1{\relax\ifmmode #1\else $#1$\fi}
\def\ls#1{\ifmath{_{\lower1.5pt\hbox{$\scriptstyle #1$}}}}

\def\tanb{\tan\beta}

\def\hl{h^0}
\def\ha{A^0}
\def\hh{H^0}
\def\hpm{H^\pm}
\def\hsm{h^0_{\scriptscriptstyle\rm SM}}
\def\mhsm{m_{h^0_{\rm SM}}}
\def\mha{m_{\ha}}
\def\mhl{m_{\hl}}
\def\mhh{m_{\hh}}

\def\mz{m_Z}
\def\mw{m_W}

\def\anti{\overline}

\def\fbi{~{\rm fb}^{-1}}

\def\gev{\,{\rm GeV}}

\def\br{{\rm BR}}
\def\rts{\sqrt s}
\def\gamhsm{\Gamma_{\hsm}^{\rm tot}}
\def\gam{\gamma}
\def\wstar{W^{\star}}

\begin{document}
\pagestyle{empty}
\begin{flushright}
SCIPP 97/07 \\
hep--ph/9703381 \\
\end{flushright}
\vskip3cm
\begin{center}
{\large\bf FUTURE DIRECTIONS IN HIGGS PHENOMENOLOGY}\\[2pc]
{ HOWARD E. HABER}\\
{\it Santa Cruz Institute for Particle Physics \\
University of California, Santa Cruz, CA 95064}
\end{center}

\vfill
\abstracts%
{The search for the weakly-coupled Higgs sector at future colliders
consists of three phases: discovery of a
Higgs candidate, verification of the Higgs interpretation of the
signal, and precision measurements of Higgs sector
properties.  The discovery of one Higgs boson with Standard Model
properties is not sufficient to expose the underlying structure of the
electroweak symmetry breaking dynamics.  It is critical to search
for evidence for a non-minimal Higgs sector and/or new physics
associated with electroweak symmetry breaking dynamics.}
\vfill
\begin{center}
Invited talk given at the \\
{\sl Ringberg Workshop on  the Higgs puzzle--- \\
What can we learn from LEP2, LHC, NLC, and FMC}\\
Ringberg Castle, Tegernsee, Germany, 8--13 December 1996.
\vfill
\end{center}
\clearpage
\setcounter{page}{1}
\pagestyle{plain}
\begin{center}
{\bf FUTURE DIRECTIONS IN HIGGS PHENOMENOLOGY}\\[6pt]
{\small HOWARD E. HABER}\\
{\small\it Santa Cruz Institute for Particle
Physics, University of California, Santa Cruz, CA 95064}
\end{center}

\vskip1pc
\abstracts%
{The search for the weakly-coupled Higgs sector at future colliders
consists of three phases: discovery of a
Higgs candidate, verification of the Higgs interpretation of the
signal, and precision measurements of Higgs sector
properties.  The discovery of one Higgs boson with Standard Model
properties is not sufficient to expose the underlying structure of the
electroweak symmetry breaking dynamics.  It is critical to search
for evidence for a non-minimal Higgs sector and/or new physics
associated with electroweak symmetry breaking dynamics.}

\section{Introduction}

The discovery of the Higgs boson would begin to
address the outstanding problem of elementary particle physics: what is
the origin of electroweak symmetry breaking and the nature of the
dynamics responsible for it.
Higgs hunting at future colliders will consist of three phases.  Phase
one is the initial Higgs boson search in which a Higgs signal is found
and confirmed as evidence for new phenomena not described by
Standard Model background.  Phase two will address the
question: should the signal be identified with Higgs physics?  Finally,
phase three will consist of a detailed probe of the Higgs sector and
precise measurements of Higgs sector observables.
Discovery of a Higgs-like signal
alone may not be sufficient to earn a place in the Particle Data Group
tables.  Some basic measurements of the properties of the Higgs
candidate will be essential to confirm a Higgs interpretation of the
discovery.

It is not unlikely that the first Higgs state to be discovered will be
experimentally indistinguishable from the Standard Model
Higgs boson ($\hsm$).
This occurs in many theoretical models that exhibit the decoupling
of heavy scalar states \cite{habernir,DECP}.  In this decoupling limit,
the lightest Higgs state, $\hl$ is a neutral CP-even
scalar with properties nearly identical to the $\hsm$, while the other
Higgs bosons of the non-minimal Higgs sector are heavy (compared
to the $Z$) and are approximately mass-degenerate.  Thus, discovery of
$\hl\simeq\hsm$ may shed little light on the dynamics underlying
electroweak symmetry breaking.
Precision measurements are critical in order to distinguish
between $\hl$ and $\hsm$ and/or to map out the properties of the
non-minimal Higgs states.

Higgs phenomenonolgy at future colliders was recently re-evaluated at
the 1996 Snowmass Workshop.  This paper reviews some of the principal
findings of the Higgs boson working group study.  Further
details can be found in Refs.~\cite{snowmass} and \cite{gunreport}.

\section{Phase 1 -- Demonstrate the Observability
of a Higgs Signal}

In the planning of future collider facilities, the machine and detector
characteristics must be developed in such a way that a Higgs signal can
be unambiguously detected above the Standard Model background.
In this paper, I shall focus on the Standard Model Higgs boson,
$\hsm$, and the Higgs bosons of the minimal supersymmetric extension of
the Standard Model (MSSM): $\hl$, $\hh$, $\ha$, and $\hpm$.
In the decoupling limit, the discovery reach of $\hsm$ at future colliders
also applies to the
lightest CP-even neutral Higgs boson ($\hl$) of the MSSM.

\begin{table}[htb]
\small
\tcaption{The $\protect \hsm$ discovery reach of future colliders.
A $5\sigma$ signal above background is required for discovery.
Note that Run II at the Tevatron complements the
LEP Higgs search only for an integrated luminosity well beyond one
year at the design luminosity of the Main Injector.
For NLC, both $\protect \sqrt{s}=500$~GeV and 1~TeV cases are shown.
The discovery reach of a $\mu^+\mu^-$ collider (FMC)
is similar to that of the NLC for the same
center-of-mass energy and integrated luminosity.}
\label{hsmdiscovery}
\renewcommand\arraystretch{1.2}
\setlength{\tabcolsep}{2pc}
\begin{center}
\begin{tabular}{lcc}
\hline \hline
                            & Integrated & Discovery \\[-2pt]
\multicolumn{1}{c}{Collider}& Luminosity &  Reach \\
\hline
LEP-2 ($\sqrt{s}=192$~GeV)&    150 pb$^{-1}$ & 95 GeV    \\
Tevatron & 5--10 fb$^{-1}$  & 80--100 GeV \\
TeV-33 & 25--30 fb$^{-1}$ & 120 GeV \\
LHC&     100 fb$^{-1}$ & 800 GeV\\
NLC-500 & 50 fb$^{-1}$ & 350 GeV \\
NLC-1000 & 200 fb$^{-1}$ & 800 GeV \\[2pt]
\hline\hline
\end{tabular}
\end{center}
\end{table}

\noindent\qquad
{\it 1. The Standard Model Higgs Boson}\\

The $\hsm$ discovery reach of future colliders is summarized above in
Table~\ref{hsmdiscovery}.  At LEP-2 running at $\sqrt{s}=192$~GeV, the
discovery reach of $\mhsm\simeq 95$~GeV can be attained by one detector
taking
data for about one year at design luminosity \cite{janot}.  With four
LEP detectors running, the Higgs mass discovery reach can be achieved
sooner (or improve on the significance of any candidate Higgs signal).
Additional luminosity cannot significantly extend the Higgs mass reach
unless the LEP-2 center-of-mass energy were increased.  At
Run II of the Tevatron one year of data taking at the Main Injector
design luminosity (1--$2~{\rm fb}^{-1}$) is
not sufficient to discover a Standard Model Higgs boson above
background.  However, two detectors running at design luminosity from
three to five years can complement the LEP-2 Higgs search.  In
particular, the associated production of $W\hsm$ with $\hsm\to b\bar b$
may be feasible at the Tevatron, given sufficient integrated
luminosity.  Assuming a
total integrated luminosity of 5~[10]~fb$^{-1}$, a Standard Model Higgs
mass discovery reach of 80~[100]~GeV is attainable \cite{tevreport,kky}.
The Tevatron Higgs search technique also applies at higher luminosity.
For example, initial studies indicate that at TeV-33, a
Standard Model Higgs boson with a mass of 120 GeV can be discovered
with an integrated luminosity of 25--30 fb$^{-1}$ \cite{tevreport,kky}.
The significance of the Higgs signal could be enhanced by the detection
of the associated production of $Z\hsm$, $\hsm\to b\bar b$ \cite{yao}.
Implicit in these studies is the assumption that the Standard
Model contributions are sufficiently well understood that the Higgs
signal can be detected as a small excess above background.

The LHC is required if one wants to extend the Higgs mass discovery
reach significantly beyond ${\cal O}(m_Z)$ \cite{gunreport,atlas,cms}.
For $\mhsm\gsim 2\mz$, the ``gold-plated mode'' $\hsm\to
ZZ\to\ell^+\ell^-\ell^+\ell^-$ provides a nearly background free
signature for Higgs boson production until the production rate becomes
too small near the upper end of the weakly-coupled Higgs mass regime.
In this case, other signatures ({\it e.g.}, $\hsm\to ZZ\to
\ell^+\ell^-\nu\bar\nu$ and $\hsm\to W^+W^-\to \ell\nu+{\rm jets}$)
provide additional signatures for Higgs discovery.
The most troublesome Higgs mass range for hadron colliders is the
so-called
``intermediate Higgs mass regime'', which corresponds roughly to
$\mz\lsim\mhsm\lsim 2\mz$.  For 130~GeV~$\lsim\mhsm\lsim 2\mz$, one can
still make use of the gold plated mode at the LHC,
$\hsm\to ZZ^{*}\to\ell^+\ell^-\ell^+\ell^-$ (where $Z^{*}$ is virtual).
Standard
Model backgrounds begin to be problematical when the branching ratio
${\rm BR}(\hsm\to ZZ^{*})$ becomes too small.  This occurs for
$2\mw\lsim\mhsm\lsim 2\mz$ where ${\rm BR}(\hsm\to W^+W^-)$ is
by far the dominant Higgs decay channel, and for
$\mhsm\lsim 140$~GeV where the the virtuality of $Z^{*}$ begins to
significantly reduce the $\hsm\to ZZ^{*}$ decay rate.
A complementary channel $\hsm\to WW^{(\ast)}\to \ell^+\nu\ell^-\bar\nu$
provides a viable Higgs signature for 155~GeV$\lsim\mhsm\lsim 2\mz$
\cite{Dreiner}, and closes a potential hole near the upper end of the
intermediate Higgs mass range.  For $\mhsm\lsim
130$~GeV, the dominant decay channel $\hsm\to b\bar b$ has very large
Standard Model two-jet backgrounds. Thus, in this regime, it is
necessary to consider rarer production and decay modes with more
distinguishing characteristics.  Among the signatures studied in the
literature are:
(i) $gg\to\hsm\to\gamma\gamma$,
(ii) $q\bar q\to V^{*} \to V\hsm\,$ ($V=W$ or $Z$),
(iii) $gg\to t\bar t\hsm$,
(iv)~$gg\to b\bar b\hsm$, and
(v) $gg\to\hsm\to\tau^+\tau^-$.
The LHC detectors are being optimized
in order to be able to discover an intermediate mass Higgs boson via its
rare $\gamma\gamma$ decay mode (with a branching ratio of about
$10^{-3}$).  The other signatures could be used to
provide consistency checks for the Higgs discovery as well as provide
additional evidence for the expected Higgs-like properties of the Higgs
boson candidate.  A successful intermediate mass Higgs search
via the $\gamma\gamma$ decay mode
at the LHC will require maximal luminosity and
a very fine electromagnetic calorimeter resolution
(at about the 1\% level).

In contrast to the Tevatron and LHC Higgs searches, the
Standard Model Higgs search at the NLC in the intermediate mass regime
is straightforward, due to the simplicity of the Higgs
signals, and the relative ease in controlling the Standard Model
backgrounds. Higgs production is detected at the NLC via two main
signatures.  The first involves the extension of the LEP-2 search for
$e^+e^-\to Z\hsm$
to higher energies.
In addition, a second process can also be significant:
the (virtual) $W^+W^-$ fusion process,
$e^+e^- \to\nu\bar\nu W^* W^* \to\nu\bar\nu\hsm$.
The fusion cross-section grows logarithmically with the center-of-mass
energy and becomes the dominant Higgs production
process at large $\sqrt{s}/\mhsm$. For example, at
$\sqrt{s}=500$~GeV, complete coverage
of the intermediate Higgs mass regime below $\mhsm\lsim 2\mz$ requires
only 5~fb$^{-1}$ of data.  The only limitation of the NLC in the Higgs
search is the center-of-mass energy of the machine which
determines the upper limit of the Higgs boson discovery reach.
One would need $\sqrt{s}\simeq 1$~TeV
to fully cover the weakly-coupled Standard Model Higgs
mass range \cite{finland,nlcreport,desyreport}.

The techniques for the Standard Model Higgs boson {\it discovery} at a
$\mu^+\mu^-$ collider (FMC) are, in principle,
identical to those employed at the
NLC \cite{BBGH,mumureport}.  However,
one must demonstrate that the extra background resulting from an
environment of decaying muons can be tamed.  It is believed that
sufficient background rejection can be achieved \cite{miller}; thus
the FMC has the same discovery reach as the NLC at the same
center-of-mass energy and luminosity. \\


\noindent\qquad
{\it 2. Higgs Bosons of the MSSM}\\

Next, we turn to the discovery potential at future colliders for the
Higgs bosons of the MSSM.  If $\mha\gg\mz$, then the decoupling limit
applies, and the couplings of $\hl$ to Standard Model particles are
identical to those of $\hsm$.  Thus, unless $\hl$ decays appreciably to
light supersymmetric particles, the discussion given above for $\hsm$
apply without change to $\hl$.  In general, one can consider two types
of MSSM Higgs searches at future colliders.  First, one can map out the
region of MSSM parameter space where at least one MSSM Higgs
boson can be discovered in a future collider Higgs search.  If no Higgs
state is discovered, then the corresponding region of MSSM parameter
space would be excluded.
(In some cases, the
absence of a Higgs discovery would be strong enough to completely rule
out the MSSM!)
Note that in this approach, one may simply discover one
Higgs state---the light CP-even neutral $\hl$---with properties
resembling that of $\hsm$, which would be consistent
with MSSM expectations, but would provide no direct proof that
low-energy
supersymmetry underlies the Higgs sector dynamics.
Second, one can examine the discovery potential for
specific states of the non-minimal Higgs sector.
In the decoupling limit, the non-minimal Higgs states are
heavy (compared to the $Z$), nearly degenerate in mass, and
weakly-coupled.  Discovery of these states at future colliders is far
from being assured.

\begin{table}[htb]
\tcaption{MSSM Higgs boson discovery potential}
\label{mssmdiscpotential}
\renewcommand\arraystretch{1.2}
\setlength{\tabcolsep}{9pt}
\begin{center}
\small
\begin{tabular}{lp{12.25cm}}
\hline\hline
\multicolumn{1}{c}{Collider}&
\multicolumn{1}{c}{Comments}\\[2pt]
\hline
LEP-2& Significant but not complete coverage, via
        $\epem\to H^+H^-$,
        $\epem\to Zh^0$ and
       $\epem\to h^0A^0$. \\[3pt]
TeV-33& Limited coverage, complements the
          LEP-2 search. \\[3pt]
LHC&  (Nearly) complete coverage for the discovery of at
      least one Higgs boson of the MSSM.
      Main challenge:  the intermediate Higgs mass region
      [$m_Z \lsim m_{h^0} \lsim 2m_Z$] which requires different
      search strategies depending on the value of $m_{h^0}$.
      Some sensitivity to heavier non-minimal Higgs states.\\[3pt]
NLC&   Complete coverage for the discovery of at least one Higgs boson of the \\ [-2pt]
and&   MSSM.  Sensitivity to heavier non-minimal states depends
        on $\sqrt s$: \\
FMC&
       \quad   $\sqrt s\gsim 2m_A$ \quad
            for discovery of $H^\pm,H^0,A^0$ via
                                        associated production. \\
&    \quad$\sqrt s\sim  m_A \9\quad$
            for $\mu^+\mu^- \to H^0,A^0$ $s$-channel
                                       resonance production.   \\[2pt]
\hline \hline
\end{tabular}
\end{center}
\end{table}

We summarize the MSSM Higgs boson discovery potential at future colliders
in Table~\ref{mssmdiscpotential}.
Consider first the discovery limits for $\hl$ of the MSSM.
The tree-level MSSM predicts that $\mhl\leq\mz$ \cite{HUNTERS}.
Suppose that this predicted bound
were unmodified (or reduced) after taking radiative corrections into
account.  Then the non-observation of $\hl$ at LEP-2 (which will
eventually be sensitive to the mass range $\mhl\lsim
95$~GeV) would rule out the MSSM.  However, for some choices of MSSM
parameters, the radiative corrections significantly {\it increase} the
tree-level bound.  Based on the most recent analyses of Ref.~\cite{carena},
if superpartner masses are no heavier than a few TeV,
then the Higgs mass bound in the MSSM is $\mhl\lsim 130$~GeV.
Consequently, the absence of a Higgs discovery
at LEP-2 and the Tevatron cannot completely rule out the MSSM.

On the other hand, it would appear that the LHC has
access to the full MSSM Higgs sector parameter space.  After all, we
noted above that the LHC will be able to completely cover the
intermediate Standard Model Higgs mass regime.  However, when
$\mha\sim{\cal O}(\mz)$,
the decoupling limit does not apply, and the properties of $\hl$ deviate
from those of $\hsm$.  Thus, an independent analysis is required
to ascertain the discovery potential of the LHC search for MSSM Higgs
bosons.  In particular, the LHC detector collaborations must demonstrate
the feasibility of $\hl$ discovery in the mass range $\mz\lsim\mhl\lsim
130$~GeV.  This is precisely the most difficult region for the LHC Higgs
search.  At this time, one can argue that the LHC coverage of the MSSM
Higgs sector parameter space is nearly complete, although the
search strategies sometimes depend on the observation of small signals (above
significant Standard Model backgrounds) in more than one channel.
Moreover, the present estimates of the statistical significance of the
Higgs signal rely on theoretical determinations of both signal and
background rates as well as simulations of detector performance.  Thus,
if no Higgs signal is confirmed by the LHC, it might still be difficult
to definitively rule out the MSSM.

The NLC (and FMC) provide complete
coverage of the MSSM Higgs sector parameter space once the
center-of-mass energy is above 300~GeV.  In contrast to the LHC
Higgs search, the intermediate Higgs mass regime presents no particular
difficulty for the high energy lepton colliders.  The associated
production
$e^+e^-\to\hl\ha$
provides an addition discovery channel for $\mha\lsim\sqrt{s}/2$.
If no Higgs signal is seen, then the
lepton colliders can unambiguously rule out the MSSM.

If only one Higgs boson is discovered, it may closely resemble the
$\hsm$.  In this case, one must address the detectability of the
non-minimal Higgs states ($\hh, \ha, \hpm, \cdots$) at future
colliders.
Detection of heavy non-minimal Higgs states at the LHC is difficult
due to the very low signal-to-background ratio of the
corresponding Higgs boson signals. In
particular, heavy Higgs states couple very weakly to gauge bosons, and
would have to be detected via their heavy fermion decays.  At large
$\tan\beta$, where the Higgs couplings to down-type fermions is
enhanced relative to the Standard Model, it may be possible to observe a
heavy neutral Higgs boson via its decay to $\tau^+\tau^-$.
At the NLC, the
main obstacle for the discovery of non-minimal Higgs states
is the limit of the center-of-mass
energy.  The heavy Higgs states of the MSSM can be produced in
sufficient number and detected only if $\sqrt{s}\gsim 2\mha$
\cite{DECP}. The
discovery reach could in principle be somewhat extended by employing the
$\gamma\gamma$ collider mode of the NLC.
In this mode of operation, the search for $\gamma\gamma\to\ha$ and
$\gamma\gamma\to\hh$ can extend the non-minimal Higgs mass
discovery reach of the NLC \cite{gunhab}.

Finally,
the FMC can produce the neutral Higgs states singly via $s$-channel
$\mu^+\mu^-$ annihilation, and would permit the discovery of the
heavy neutral Higgs states up to $\sqrt{s}=\mha$ \cite{BBGH}.
The viability of this discovery mode depends on the parameters of the
Higgs sector.  In the MSSM, the cross-section for
$\mu^+\mu^-\to \hh,\ha$
is enhanced for values of $\tan\beta$ above 1.  For $\mhh, \mha\gg\mz$,
$\hh$ and $\ha$ are approximately degenerate in mass.
Given sufficient luminosity, one can detect $\hh$ and
$\ha$ (if kinematically accessible) by scanning in $\sqrt{s}$, assuming
that $\tan\beta$ is larger than a critical value (which depends on the
total luminosity and the Higgs mass).  Detection is accomplished via a
resonant peak in the Higgs decay to $b\bar b$ (and $t\bar t$ if
allowed).

\section{Phase 2 -- After Discovery: Is It a Higgs Boson?}

Suppose that the first candidate Higgs signal is detected.
What must one do to prove that the produced state is a Higgs boson?
We assume that after the initial discovery is made, further collider
running confirms the signal and establishes a useful statistical sample
of events.  A list of the primary Higgs signals at future colliders
is given in Table~\ref{tab:hsmsignatures}.

\begin{table}[htb]
\tcaption{Primary $\hsm$ signatures at future colliders and the
corresponding Higgs mass range over which detection of a
statistically significant signal is possible.}
\label{tab:hsmsignatures}
\renewcommand\arraystretch{1.5}
\setlength{\tabcolsep}{2pc}
 \begin{center}
 \small
\begin{tabular}{lll}
\hline\hline
\multicolumn{1}{c}{Collider}&
\multicolumn{1}{c}{Signature}&
\multicolumn{1}{c}{Mass Range}\\[2pt]
\hline
LEP-2&    $e^+e^-\to Z\hsm$&   $\lsim 95$ GeV \\
TeV-33$^a$& $ W^*\to W\hsm\to \ell\nu b\bar b$&   60--120 GeV \\
  & $Z^*\to Z\hsm\to \left\{\matrix{\lplm b\bar b \cr\noalign{\vskip3pt}
                          \nu\bar\nu\ b\bar b\quad}\right.$
                                            &  \\[3pt]
LHC&  $W^*\to W\hsm\to \ell\nu b\bar b$& 80--100 GeV \\
   & $\hsm + X\to \gamma\gamma+X$& 90--140 GeV \\
   & $\hsm\to ZZ^* \to\lplm\lplm$& 130--180 GeV \\
   & $\hsm\to WW^*\to \ell^+\nu\ell^-\bar\nu$& 155--180 GeV \\
   & $\hsm\to ZZ\to \lplm\lplm$& 180--700 GeV \\
   & $\hsm\to ZZ \to \nu\bar\nu\lplm$& 600--800 GeV \\
   & $\hsm\to W^+W^- \to \ell\nu\ +$\ jets& 600--1000 GeV
\protect\cite{luc}\\[3pt]
$\matrix{\rm NLC \cr \hphantom{NLC}\cr \hphantom{NLC} }$&
$\left.\matrix{ \epem \to Z\hsm\quad\cr\noalign{\vskip3pt}
        \epem\to \nu\bar\nu\hsm\quad \cr\noalign{\vskip3pt}
        \epem\to \epem\hsm} \right\}$& $\lsim 0.7 \sqrt s$ \\[1.6pc]
$\matrix{\rm FMC\cr \hphantom{FMC}\cr \hphantom{FMC} }$&
$ \left.\matrix{ \mpmm \to Z\hsm \quad\;\cr \noalign{\vskip3pt}
          \mpmm\to \nu\bar\nu\hsm \quad\;\cr  \noalign{\vskip3pt}
          \mpmm\to \mpmm\hsm } \right\}$& $\lsim 0.7 \sqrt s$ \\[2pt]
   & $\mpmm\to\hsm$                  &up to $\sqrt s<2\mw$ \\ [3pt]
\hline \hline
\end{tabular}
\end{center}
{\footnotesize
\vskip3pt
\hangafter{$^a$}The TeV-33 Higgs signatures
listed above are also relevant for lower luminosity Tevatron searches
over a more restricted range of Higgs masses, as indicated in
Table~\ref{hsmdiscovery}.

}
\end{table}

The first step is to ascertain whether
the observed state resembles the Standard Model Higgs boson and/or if
it is associated with a non-minimal Higgs
sector.  If $\hl\simeq\hsm$, then one must demonstrate that the
discovered state has
(i) zero electric and color charge,
(ii) spin zero,
(iii) CP-even quantum number,
(iv)  electroweak strength couplings, and
(v)   couplings proportional to the mass of the state to which it
couples.
Eventually, one would like to make detailed
measurements and
verify that the Higgs candidate matches all the properties expected of
$\hsm$ to within some precision (small deviations from the $\hsm$
properties
will be addressed in Phase 3).  If the properties of the
discovered state are Higgs-like, but differ in detail from those of
$\hsm$, then it is likely that other non-minimal Higgs states are light
and may have been produced in the same experiment.  Finding evidence for
these states will be crucial in verifying the Higgs interpretation of
the data.

At an $e^+e^-$ collider (LEP-2 and the NLC),
many of the Higgs boson properties can be directly measured due to
low backgrounds and simple event structures.\footnote{In principle,
the remarks that follow also apply to the FMC.  However, it has not yet
been demonstrated that the severe backgrounds arising from the
constantly decaying muons can be overcome to make precision
measurements.}
One can directly
measure the spin and CP-quantum numbers of the Higgs candidate through
the angular distributions of production and decay.  Specific Higgs decay
modes can be separated and individually studied.  Accurate measurements
of $\sigma(\hl){\rm BR}(\hl\to X)$ can be made for a number of final
states, including $X=b\bar b$ and $\tau^+\tau^-$.  A recent
breakthrough was made which demonstrates that detection of $\hl\to c\bar
c$ is possible with appreciable efficiency and low mis-identification
\cite{bctags}. Thus, at the lepton colliders, $\hl\simeq\hsm$ can be
confirmed with some precision.

The verification of a Higgs interpretation of
a Higgs signal discovered at a hadron collider is much
more involved.
One must examine in detail a variety of
possible Higgs signatures
(see Table~\ref{tab:hsmsignatures}) and evaluate the potential
of each channel for supporting the Higgs interpretation of the signal.
Taken one by one, each channel provides limited information.
However, taken together, such an analysis might provide a strong
confirmation of the Higgs-like properties of the observed state as well
as providing a phenomenological profile that could be compared to the
predicted properties of the Standard Model Higgs boson.

The quantum numbers of the Higgs candidate may be difficult to measure
directly at a hadron collider.
However, note that if $\hsm\to\gamma\gamma$ is seen, then the $\hsm$
cannot be spin-1 (by Yang's theorem).  This does not prove that $\hsm$
is spin-zero, although it would clearly be the most likely possibility.
If the coupling $\hsm VV$ is seen at a tree-level
strength, then this would confirm the presence of a CP-even component.
Unfortunately, any CP-odd component of the
scalar state couples to $VV$ at the
loop level, so one would not be able to rule out {\it a priori} a
significant CP-odd component for $\hsm$.

The most problematical Higgs
mass range is 100~GeV$\lsim\mhsm\lsim 130$~GeV.  Higgs bosons in this
mass range are not accessible to LEP-2 or Run II of the Tevatron.  At
the LHC, the most viable signatures in this mass range involve the
production of $\hsm$
followed by $\hsm\to\gamma\gamma$.  However, the Higgs can be produced
via a number of different possible mechanisms:
(i)   $gg\to\hsm$,
(ii)  $q\bar q\to q\bar q\hsm \;$
            via $t$-channel $W^+W^-$ fusion,
(iv)  $q\bar q\to V\hsm\; $ via
         $s$-channel $V$-exchange, and
(v)    $gg\to t\bar t\hsm$.
The $gg\to\hsm$ mechanism dominates, and it will be an experimental
challenge to separate out the other production mechanisms.
It may be possible to separate $gg\to\hsm$ and $W^+W^-\to\hsm$ events
using a forward jet tag which would select out the $W^+W^-$ fusion
events.  It may also be possible to distinguish $V\hsm$
($V=W^\pm$ or $Z$) and $t\bar
t\hsm$ events based on their event topologies.  If these other
production mechanisms can be identified,
then it would be possible to extract information about
relative couplings of the Higgs candidate to $VV$ and $t\bar t$.
Otherwise, one will be
forced to rely on matching $\sigma(\hsm){\rm BR}(\hsm\to\gamma\gamma)$
to Standard Model expectations in order to confirm the Higgs
interpretation of $\hsm$.

In some circumstances, it might be possible to observe the decays
$\hsm\to b\bar b$ or $\hsm\to\tau^+\tau^-$ (after a formidable
background subtraction), or identify the Higgs boson produced via $gg\to
b\bar b\hsm$.  One could then extract the relative coupling strengths
of $\hsm$ to $b\bar b$ and/or $\tau^+\tau^-$ final states.  These could
be compared with the corresponding $VV$ and $t\bar t$ couplings,
and confirm that the
Higgs candidate couples to particles with coupling strengths proportional
to the particle masses.

As a result of these considerations, Ref.~\cite{gunreport} concludes
that in some Higgs parameter range, LHC can
make a convincing case for the ``expected''
Higgs-like properties of a Higgs signal.
Ratios of Higgs couplings to different final
states may be measured to roughly 20--30\%.


%
\section{Phase 3 -- Precision Measurements of Higgs Properties}

Let us suppose that the Higgs candidate (with a mass no larger than a
few times the $Z$ mass)  has been confirmed to have the
properties expected of the $\hsm$ (to within the experimental error).
One would then be fairly confident that the
dynamics that is responsible for electroweak symmetry breaking is
weakly-coupled.
Unfortunately, the details of the underlying physics responsible for
electroweak symmetry breaking would still be missing.
As a consequence of the decoupling of heavy Higgs states, it is
possible to construct many models of
scalar dynamics that produce a light scalar state with the properties
of the $\hsm$.  To distinguish among such models,
additional properties of the scalar sector must be
uncovered. It is the non-minimal Higgs states that encode the structure
of the electroweak symmetry breaking dynamics.
In order to provide experimental proof of the existence of a non-minimal
Higgs sector, one must either demonstrate that the properties of $\hl$
differ (even if by a small amount) from those of $\hsm$, or one must
directly produce and detect the heavier Higgs states ($\hh, \ha,
\hpm, \cdots$).  In general, precision measurements of both light
and heavy Higgs properties
are essential for distinguishing among models of electroweak symmetry
breaking dynamics.

\begin{table}[h]
\tcaption{Anticipated experimental errors in the measured values of the
$\hsm$  branching ratios, the partial decay rate, $\Gamma(\hsm\to\gamma
\gamma)$, and total width, $\gamhsm$, in percent, for
various ranges of $\mhsm$.  The notation ``?'' indicates that a
reliable simulation or estimate is not yet available or that the number
indicated is a very rough guess, while ``--''
means that the corresponding observable cannot be reliably measured.
The results listed below are primarily derived from a multi-year run at
the NLC.  For $\hsm\to\gamma\gamma$, data from LHC and the
$\gamma\gamma$ collider are also employed to improve the quoted errors.
The total Higgs decay rate can be obtained indirectly (by combining
measurements of related quantities); the comparison with the direct
determination via $s$-channel Higgs resonance production at the FMC is
shown.  See the text and Ref.~\protect\cite{gunreport} for further
details.}
\renewcommand\arraystretch{1.5}
\setlength{\tabcolsep}{1.5pc}
\begin{center}
\small
\begin{tabular}{lcccc}
\hline
\hline
 &  \multicolumn{4}{c}{$\mhsm$ range (GeV)} \\[-3pt]
\multicolumn{1}{c}{Observable}&
              80--130 &  130--150 &  150--170 & 170--300\\[3pt]
\hline \noalign{\vskip3pt}
 $\br(\hsm\to b\anti b)$& 5--6\% & 6--9\% & 20\% ? & $-$ \\
 $\br(\hsm\to c\anti c)$& $\sim 9\%$ & ? & ? & $-$ \\
 $\br(\hsm\to W\wstar)$\hspace*{-9pt} & $-$ & 16--6\% &6--5\% & 5--14\%
\\
 $\br(\hsm\to\gam\gam)$ & 15\% & 20--40\% & ? & $-$  \\
 $\Gamma(\hsm\to\gam\gam)$ & 12--15\% & 15--31\% & ? & 13--22\% \\
 $\gamhsm$ (indirect) & 19--13\% & 13--10\% & 10--11\% &  11--28\% \\
 $\gamhsm$ (FMC) & 3\%$^a$& 4--7\%& $-$& $-$\\[5pt]
\hline \hline
\end{tabular}
\end{center}
\vskip3pt
{\footnotesize \hspace*{.4cm}
$^a$Near the $Z$ peak, the expected FMC uncertainty in $\gamhsm$
 is about 30\%.
}
\label{widthbrs}
\end{table}

The precision measurements of Higgs properties include
branching ratios, cross-sections, and quantum numbers as
previously discussed.  In Phase 3,
it is important to be able to separate cross-sections and
branching ratios (instead of simply measuring the product of the two).
More
challenging will be the measurement of absolute partial widths, which
requires a determination of the total Higgs width.  
Below $ZZ$ threshold, the Standard Model Higgs width is too small to be
directly measured, and other strategies must be employed.
As an illustration, Table~\ref{widthbrs} presents the
anticipated errors in the measurements of some $\hsm$ branching
ratios, the partial decay rate for $\hsm\to\gamma\gamma$, and the total
Higgs width, $\gamhsm$, for $80\leq\mhsm\leq
300\gev$.  The quoted errors are determined primarily
by considering the data that would be collected by the NLC at
$\rts=500\gev$ with a total integrated luminosity of $L=200\fbi$.
For BR($\hsm\to\gam\gam$), the NLC analysis has been combined
with results from an LHC analysis; while the measurement of
$\Gamma(\hsm\to\gam\gam)$ relies on data taken from a 
50~fb$^{-1}$ run in the $\gamma\gamma$ collider mode of the NLC
(with an $e^+e^-$ center-of-mass energy of $\sqrt{s}\sim 1.2\mhsm$).
These quantities also contribute to the net accuracy of the
total Higgs width, $\gamhsm$, following the indirect
procedure%
\footnote{For $\mhsm\lsim 130$~GeV, the indirect procedure
relies on the $\hsm\to\gamma\gamma$ measurements.  In the case of
$\mhsm\gsim 130$~GeV, one may also
make use of the $WW\hsm$ coupling strength extracted from data.}
described in Ref.~\cite{gunreport}.
Note that $\gamhsm$ can be measured directly
only in the $s$-channel Higgs production at the FMC.  For comparison
with the indirect determination of $\gamhsm$,
the FMC scan results listed in Table~\ref{widthbrs}
assume that a total luminosity of $L=200\fbi$ is devoted to the scan.
With the exception of the case where $\mhsm\simeq\mz$, the FMC would
provide the most precise measurement of the total Higgs width for
values of the Higgs mass below the $W^+W^-$ threshold.

In models of non-minimal Higgs sectors, precision measurements of the
branching ratios and partial (and total) decay rates of the lightest
CP-even Higgs boson could prove that $\hl\neq\hsm$, thereby
providing indirect evidence of the non-minimal Higgs states.
Once the non-minimal Higgs bosons are directly discovered, detailed
measurements
of their properties would yield significant clues to the underlying
structure of electroweak symmetry breaking.  For example,
if the Higgs sector arises from a two-doublet model,
then precision studies of the heavy Higgs states can provide a
direct measurement of the important parameter $\tanb$ (the ratio of
Higgs vacuum expectation values).%
\footnote{Note that in the
decoupling limit (where $\hl$ cannot be distinguished from $\hsm$),
measurements of processes involving
$\hl$ alone cannot yield any information on the value of $\tanb$.}
The measurement of $\tanb$ can also provide a critical
self-consistency test of the MSSM, since the parameter $\tanb$ also
governs the properties of the charginos and neutralinos (and can in
principle be determined in precision measurements of supersymmetric
processes).  Moreover, the couplings of Higgs bosons to supersymmetric
particles will provide invaluable insights into both the physics of
electroweak
symmetry breaking and the structure of low-energy supersymmetry.  The
possibility that the heavy non-minimal Higgs states have non-negligible
branching ratios to supersymmetric partners can furnish an additional
experimental tool for probing the Higgs boson--supersymmetry connection.

As in the case of the
$\hsm$ discussed above, the lepton colliders (assuming that $\sqrt{s}\gsim
2\mha$ for the NLC and $\sqrt{s}\sim\mha$ for the FMC) provide the
most powerful set of tools for extracting the magnitudes of the
Higgs couplings to fermion and vector boson pairs.
The Higgs couplings
to vector boson pairs directly probe the mechanism of electroweak
symmetry breaking.
The Higgs coupling to two photons, depends (through their one-loop
contributions)
on all charged states whose masses are generated by their
couplings to the Higgs sector.  Precision
measurements of the Higgs couplings to fermions are sensitive to other
Higgs sector parameters \cite{HUNTERS}
({\it e.g.}, $\tan\beta$ and the neutral
Higgs mixing parameter $\alpha$ in a two-Higgs-doublet model).
Additional information can be ascertained if Higgs self-interactions
could be directly measured.  This would in principle provide direct
experimental access to the Higgs potential.
Unfortunately, there are
very few cases where the measurement of Higgs self-couplings has been
shown to be viable \cite{DHZ}.

\section{Conclusions}

The methods
by which the first Higgs signal will be identified are well known and
have been studied in great detail.  However, the most outstanding
challenge facing the Higgs searches at future colliders lies in
identifying and exploring in detail the properties of the Higgs states.
Precision measurements may be able to distinguish between the
Higgs boson of the Standard Model and the lightest scalar of a
non-minimal Higgs sector.
It is also crucial to directly detect
and explore the properties of the non-minimal Higgs states.
A successful
exploration will have a profound effect on our understanding of
TeV-scale physics.

\section{Acknowledgments}

This work benefited greatly from conversations with Jack Gunion.  I am
grateful for his input and collaboration.  I would like to thank Tao Han
and Michael Peskin for their careful reading of the manuscript; their
comments were extremely useful in preparing the final draft of
this review.  This work was supported in part by the U.S. Department of
Energy.

%
\end{document}